\pdfoutput=1

\documentclass{article}
\usepackage{arxiv}

\usepackage{tikz}
\usepackage{amsmath}

\usepackage{filecontents}
\usepackage{biblatex}
\bibliography{references}

\usepackage{listings}
\usepackage{multirow}

\begin{document}

\title{\Large \bf NVMe and PCIe SSD Monitoring in Hyperscale Data Centers }

\author{
Nikhil Khatri\thanks{nikhilkhatri97@gmail.com}\\
LinkedIn\\
\texttt{nikhatri@linkedin.com}
\And
Shirshendu Chakrabarti\thanks{shirshen12@gmail.com}\\
LinkedIn\\
\texttt{schakrabarti@linkedin.com}\\
}

\maketitle

\begin{abstract}
With low latency, high throughput and enterprise-grade reliability, SSDs have become the de-facto choice for storage in the data center. As a result, SSDs are used in all online data stores in LinkedIn. These apps persist and serve critical user data and have millisecond latencies. For the hosts serving these applications, SSD faults are the single largest cause of failure. Frequent SSD failures result in significant downtime for critical applications. They also generate a significant downstream RCA (Root Cause Analysis) load for systems operations teams. A lack of insight into the runtime characteristics of these drives results in limited ability to provide accurate RCAs for such issues and hinders the ability to provide credible, long term fixes to such issues. In this paper we describe the system developed at LinkedIn to facilitate the real-time monitoring of SSDs and the insights we gained into failure characteristics. We describe how we used that insight to perform predictive maintenance and present the resulting reduction of man-hours spent on maintenance.
\end{abstract}

\section{Introduction}
SSDs have gained popularity in recent times as cost per GB has reduced and performance has improved. Owing to their high throughput and low latency, they are increasingly used in on-line data stores. These applications are expected to have very low response times, coupled with high availability. Each SSD failure incurs a high cost. When a host with an SSD is removed from a cluster, it causes the cluster to re-balance. This results in an increase in query latency for the application spanning the re-balance duration, which impacts downstream services. Each failure requires the attention of site reliability engineering (SRE) teams to take the failing host out of its cluster, and the involvement of systems operations engineers to identify the root cause of failure and perform corrective action. In the event of a failure requiring SSD replacement, data-center technicians are required to physically replace the host's drive. This incurs a cost both in terms of time and the actual dollar value of the hardware. Thus, the goal of our work was to have fewer failures, and in the occurrence of a failure, be able to provide an insightful timely root cause analysis.

\section{Existing fault detection system}\label{existing}
The existing system in LinkedIn for monitoring SSD failures consists of two primary components. The first is an on-host script, that runs a series of checks to determine drive health. These are vendor-specified checks, which typically search log files for known error messages generated by the device driver, indicating a breach of internal thresholds or error events. 
These checks have, in our experience, provided inadequate indication of faults. Specifically, there are several instances of false-positives, where the vendor-specified logs were seen, without there being any application-impacting fault. Logs relating to warranty expiration and lifetime drive statistics do not provide insight into drive performance and generate noise in the ticket system. Further, there are some classes of failures for which there are no indicators provided by the manufacturer. These go entirely undetected, until the application is directly impacted.
The on-host script, on finding an error, logs this to a dedicated file, which is collected by a separate service. This service collects all such logs across hosts, and sends these to a master host which handles ticket generation.
There is a latency of up to 45 minutes between the on-host script reading the log and the ticket being generated. This is owing to the latency of log aggregation. During this time, a drive's performance may further degrade and cause the SSD to entirely stop serving IO requests. This workflow does not take into account the time taken by an engineer to pick up the ticket and notify the impacted application owners which adds another few minutes.

In summary, the key problems with the existing system are:
\begin{enumerate}
    \item Inadequate checks at host-level - Current checks are insufficient and inaccurate.
    \item High latency of alerting system - 45 minutes is a very long time for a failure alert.
\end{enumerate}

This paper focuses on improving the on-host fault detection system. For this, a deeper understanding of SSDs, their architecture, performance and failure characteristics is required. Using this knowledge, more precise checks may be written.

\section{SSDs at LinkedIn}\label{ssdsatli}
It is imperative to discuss the drives which are used in LinkedIn's data centres, as the tooling and diagnostic data available for each vary greatly. The oldest drives deployed in our datacenters are PCIe cards from Virident. The remaining are all NVMe drives from leading vendors. Table \ref{tab:dist_tab} describes the distribution of SSD types at LinkedIn and their respective failure rates. The NVMe drives in our fleet have a much lower failure rate than the non-NVMe PCIe protocol drives, and are hence not as significant a source of ticket load.

\begin{table}[ht]
    \centering

    \begin{tabular}{|c|c|c|}
        \hline
        SSD type &  Share of drives & Share of failures\\
        \hline\hline
        PCIe & 22.1\% & 77.0\%\\
        NVMe & 77.9\% & 23.0\%\\
        \hline
    \end{tabular}
    \caption{SSD and SSD failure distribution - Q2 2019}
    \label{tab:dist_tab}
\end{table}

Further aggravating the problem, is the fact that as of Q2 2019, warranties on all custom PCIe protocol cards had expired. As a result, there is virtually no vendor support for these products. SSD faults had been dealt with primarily by contacting vendor support, and the typical end result was a drive replacement, covered under the warranty. Without the warranty, replacing each faulty drive is not a cost-effective solution. Thus, a solution which extends the lifetime of drives shall have a substantial monetary impact.

\section{Health indicators}\label{indicators}
The existing system described in Section \ref{existing} relies primarily on logs written by the driver. We consider this approach inherently flawed as this relies on thresholds that are not tunable, and the rationale for which is internal to the driver, and not available for review. Thus, the new system must be based on metrics read directly from the drive, without any opaque processing by the driver. This will allow us to form our own logic to identify degradation and signal imminent failure. To develop such a system, there are two prerequisites:

\begin{enumerate}
    \item A way to extract metrics from the drive, either through diagnostic tools such as $smartctl$, or through vendor-specific tooling.
    \item A thorough understanding of what each metric means, and how it affects the reliability and performance of the drive.
\end{enumerate}

\subsection{NVMe Health Indicators}
All NVMe drives support the NVMe command set, and can be controlled using nvme-cli \cite{nvme2019}. However, this interface is not entirely uniform, since most vendors support only a subset of the NVMe commands, and add their own extensions beyond what is required by the specification. The nvme-cli tool allows access to SMART (Self-Monitoring, Analysis and Reporting Technology) attributes for SSDs. Several studies exist describing the correlation between a various SMART metrics and HDD (Hard Disk Drive) failure \cite{pinheiro2007failure, zhu2013proactive, xu2018improving}. Relatively few similar studies exist for SSDs. A detailed study of SSD failures at scale can be found in the work done at Facebook \cite{meza2015large}. This provides a good background of what metrics are of value, and are good indicators of failure. Below, we describe the selection of metrics that we collect from all NVMe SSDs.

\begin{table}[ht]
    \centering
    \begin{tabular}{|p{0.3\columnwidth}|p{0.6\columnwidth}|}
	 \hline
	 Metric & Description\\ 
	 \hline\hline
	 Flash bytes written & Bytes written to NAND flash cells    \\ 
	 \hline
	 Flash bytes read & Bytes read from NAND flash cells\\ 
	 \hline
	 Host bytes written & Amount of data written by host.\\ 
	 \hline
	 Host bytes read & Amount of data read by host\\ 
	 \hline
	 Spare flash remaining & \% of overprovisioned flash remaining\\ 
	 \hline
	 Spare flash threshold & Threshold \% for overprovisioned flash\\ 
	 \hline
	 Temperature & Drive temperature, measured on-board\\ 
	 \hline
	 CRC error count & Number of CRC checks which failed\\ 
	 \hline
	 Power cycles & Number of device power cycles\\
	 \hline
    \end{tabular}
    \caption{SMART SSD metrics}
    \label{tab:smarts}
\end{table}

In addition to the standard SMART metrics available through nvme-cli, there are some additional metrics that may be exposed through nvme-cli extensions, or in some cases, through vendor-specific tools. Some examples of vendor specific metric classes are given below.

\begin{enumerate}
    \item Wear-levelling information: This class of metrics provides summary statistics of PE (Program Erase) counts for erase blocks in the SSD. Typically min, max and average values across the SSD are available.
    \item Lifetime temperature statistics: Records of lifetime-highest, lifetime-lowest and lifetime-average temperatures are maintained. This allows engineers to determine if there are ambient problems which cause the SSD to permanently operate at higher temperatures.
    \item Program/erase fail counts: Number of times a program or erase event on a NAND block failed.
    \item Several classes of errors are recorded:
    \begin{itemize}
        \item Re-read events
        \item End to end error counts
        \item Read errors recovered without delay
        \item Read errors recovered with delay
    \end{itemize}
\end{enumerate}

\subsection{PCIe SSD Health Indicators}
For our PCIe cards, there is no standard interface. However, there is a rich toolset, which provides commands for formatting and configuring the drive, and also reading drive health related information from the controller. 
This tooling was originally meant to prepare a diagnostic package to send to the vendor in event of failure, but the data is not encrypted, and can hence be parsed. The metrics available from the vendor tooling are far more detailed than those provided by NVMe drives. These include information about physical characteristics, deeper error reporting capabilities and several metrics for each Slice of flash in the SSD. Example metrics are shown in Table \ref{tab:pcie_mets}.

\begin{table}[ht]
    \centering
    \begin{tabular}{|p{0.2\columnwidth}|p{0.2\columnwidth}|p{0.4\columnwidth}|}
        \hline
        Metric class & Metric & Description\\
        \hline\hline
        \multirow{8}{*}{Physical}& Temperature & Temperature measured by multiple on-board sensors \\
        \cline{2-3}
        & Voltage & PCIe bus voltage\\
        \cline{2-3}
        & Power & Power drawn from PCIe bus\\
        \cline{2-3}
        & Current & Current drawn from PCIe bus\\
        \hline
        \multirow{8}{*}{Error} & DPP error & Data Path Protection Error\\
        \cline{2-3}
        & RAID error & Error constructing data from internal RAID pair\\
        \cline{2-3}
        & Mark to bad error & Error while marking a flash block as bad\\
        \cline{2-3}
        & GC failure & Internal garbage collector failures\\
        \hline
        \multirow{12}{*}{per-slice} & Bytes written & Bytes written to cells in each slice\\
        \cline{2-3}
        & Bytes read & Bytes read from cells in each slice\\
        \cline{2-3}
        & Erase count & Number of erase commands for blocks in each slice\\
        \cline{2-3}
        & Bad Block count & No of blocks marked as bad for each slice\\
        \cline{2-3}
        & PE count & Max, min and average Program-Erase cycles for blocks in each cell\\
        \hline
    \end{tabular}
    \caption{PCIe SSD Metrics}
    \label{tab:pcie_mets}
\end{table}

\section{System architecture}\label{sysarch}
\begin{figure}[ht]
    \centering
    \includegraphics[width=0.9\columnwidth]{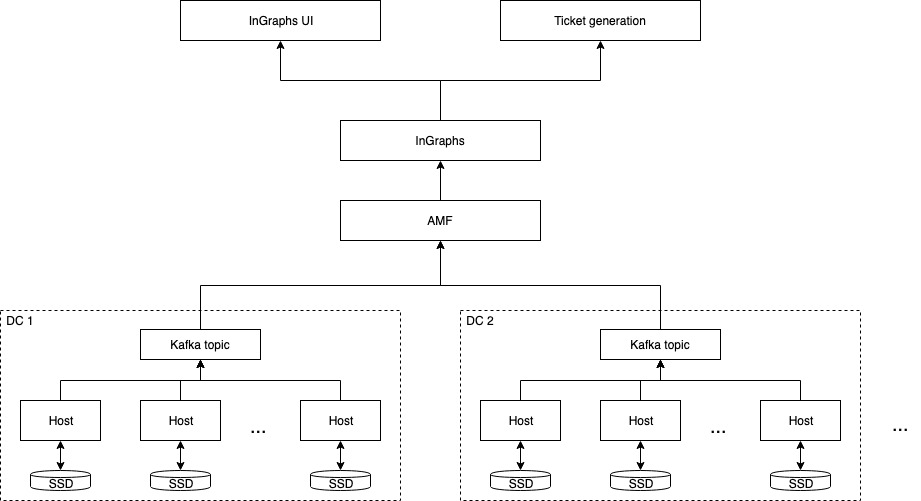}
    \caption{System architecture}
    \label{fig:hld}
\end{figure}

Section \ref{indicators} describes the on-host portion of the new solution, which solves the first problem mentioned in Section \ref{existing}, that of recording insufficient data at host-level. The second problem was the large latency between an error being detected on a host and a ticket being generated. In LinkedIn, there is separate monitoring infrastructure in place that is used by application SRE teams to monitor app-level performance metrics. This infrastructure already includes the components required to send metrics to a single source for plotting, thresholding, alerting and ticket generation. This pipeline has an end-to-end latency of up to 5 minutes, from on-host data collection, to an alert being generated. This is a significant improvement over the extant system, and is sufficiently low for our use case. Thus, we can use the existing pipeline, without having to develop a new solution. Since this infrastructure was not developed as part of the effort described in this paper, we provide a short summary of this. Figure \ref{fig:hld} provides a high level view of the monitoring pipeline employed as part of our solution.

The on-host monitoring application collects the metrics described in Section \ref{indicators} and sends these to a REST endpoint exposed by Kafka. Kafka \cite{koshy_2016} is a distributed open source streaming platform, developed at LinkedIn, and open-sourced in 2011.  Effectively, each host which contains an SSD has a Kafka producer running, writing to a dedicated topic. This topic is consumed by an application called AMF, which is the framework that ingests metrics sent by multiple services into the monitoring pipeline. This data is then passed on to InGraphs \cite{wong_2011}, a tool used to visualize time series metrics. Sending this data to InGraphs has several benefits. This platform allows engineers to have a deep insight into the characteristics of each drive, with just a few clicks. It also allows us to aggregate data based on custom classification and filtration. In addition to the immediate benefits of having a visual representation available, the InGraphs pipeline also allows us to set alerts on these graphs. These alerts are triggered by breahc of thresholds we define on certain metrics. The alerts provided by this system are very low latency, and can be configured to page the on-call teams, bypassing the ticket generation system. This ensures a faster response, specially in case of emergencies where a drive failure is imminent.

\section{Observations}\label{obs}
Having described the infrastructure that was set up to monitor SSDs in LinkedIn's datacentres, in this section we describe certain observations that were made using the newly available metrics.

\subsection{Drive temperatures}
\begin{figure}[ht]
    \centering
    \includegraphics[width=\columnwidth]{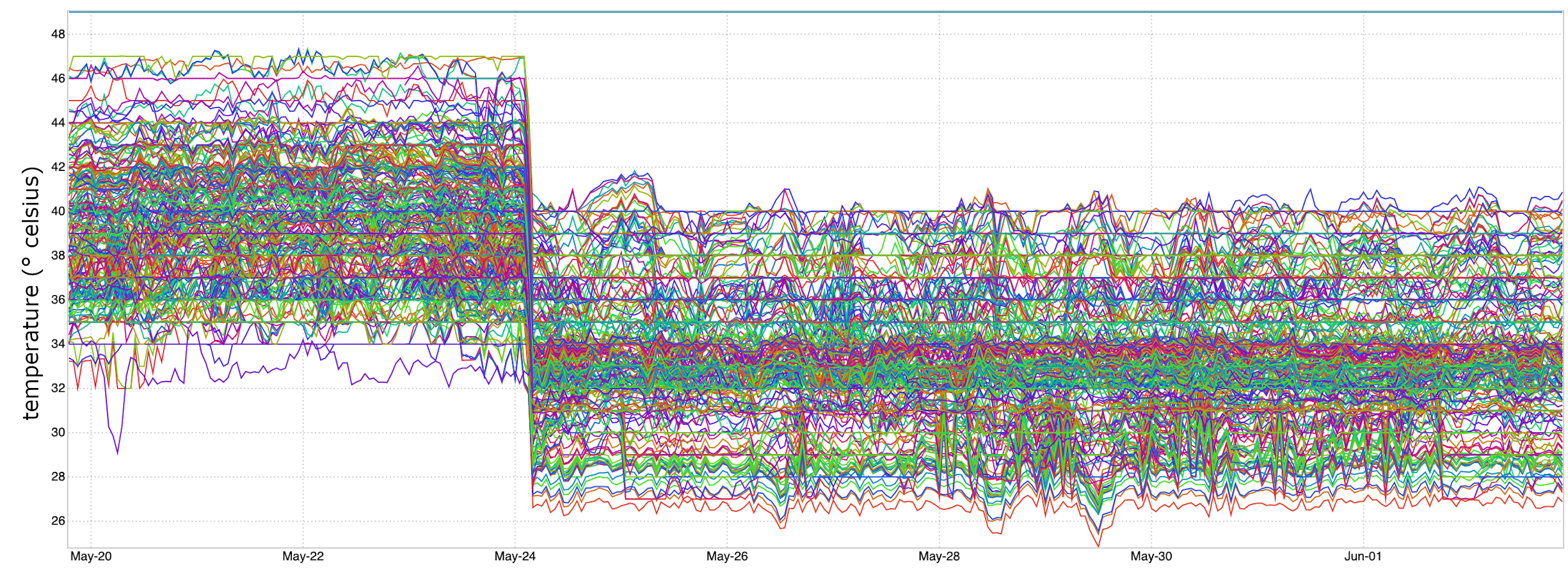}
    \caption{NVMe drive temperatures}
    \label{fig:temp_drop}
\end{figure}

\begin{figure}[ht]
    \centering
    \includegraphics[width=\columnwidth]{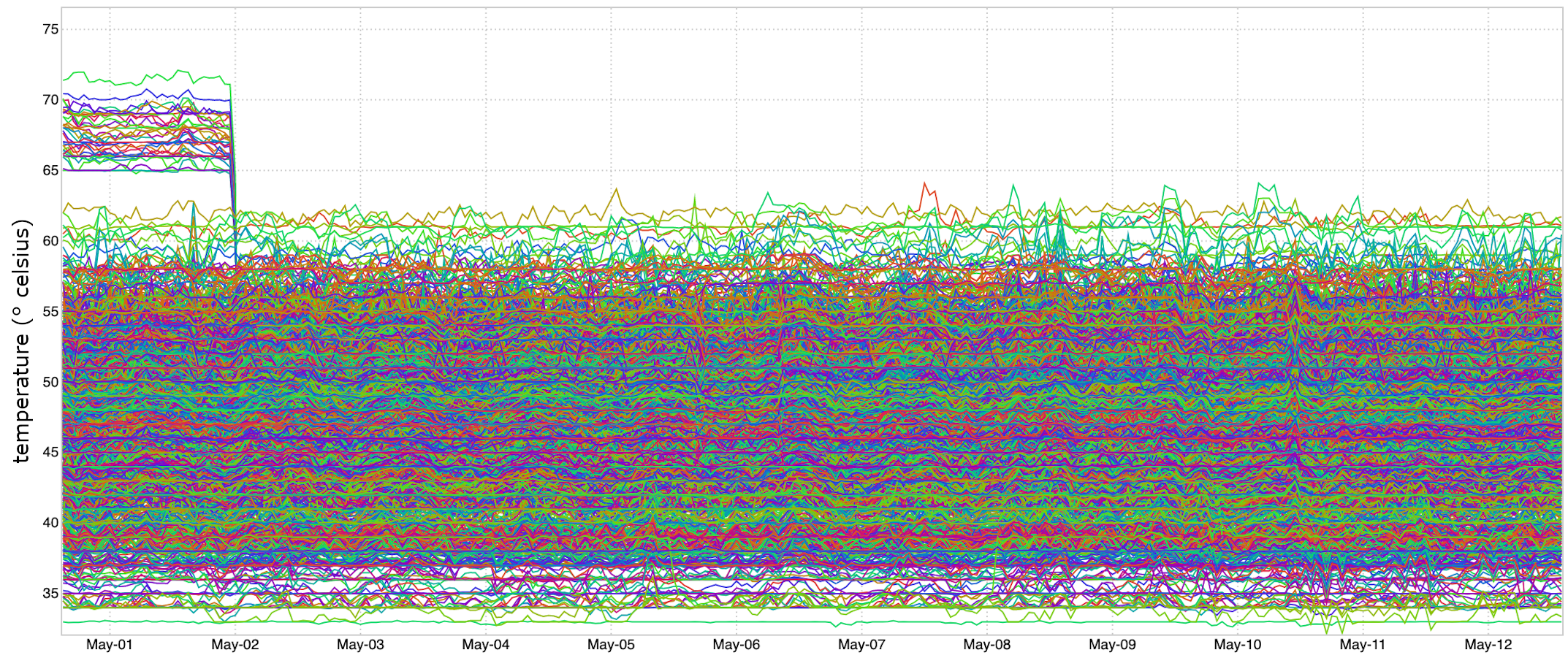}
    \caption{PCIe drive temperatures}
    \label{fig:pcie_temps}
\end{figure}

Temperature is a metric that is collected in some form from all SSD models. Thermal throttling tickets constitute a significant portion of all SSD tickets raised. On observing the graphs for SSD temperatures, it was apparent that several SSDs were running at much higher temperatures than is recommended. On further investigation, it was found that all hosts which had SSDs running at higher temperatures, were operating with an incorrect chassis fan policy. It is recommended by our vendors that we operate the fans at maximum speed for hosts which run certain SSD models. On correcting the fan policy, an immediate drop in drive temperatures was observed. This data can be seen in Figure \ref{fig:temp_drop} and Figure \ref{fig:pcie_temps}.

The data shown in Figure \ref{fig:temp_drop} represents a drop in average temperature from 40$^{\circ}$ Celsius to 33$^{\circ}$Celsius for a particular NVMe drive across all datacentres. Figure \ref{fig:pcie_temps} shows a similar drop in temperatures for PCIe SSDs in a production datacentre. Here we observe that SSD temperatures fit a bimodal distribution. These two bands corresponded exactly to the fan policy on the host. It is worth noting that the PCIe drives operate at a much higher temperature than the NVMe drives, with the hottest drive reaching temperatures upwards of 70$^{\circ}$ Celsius.

\begin{figure}[ht]
    \centering
    \includegraphics[width=\columnwidth]{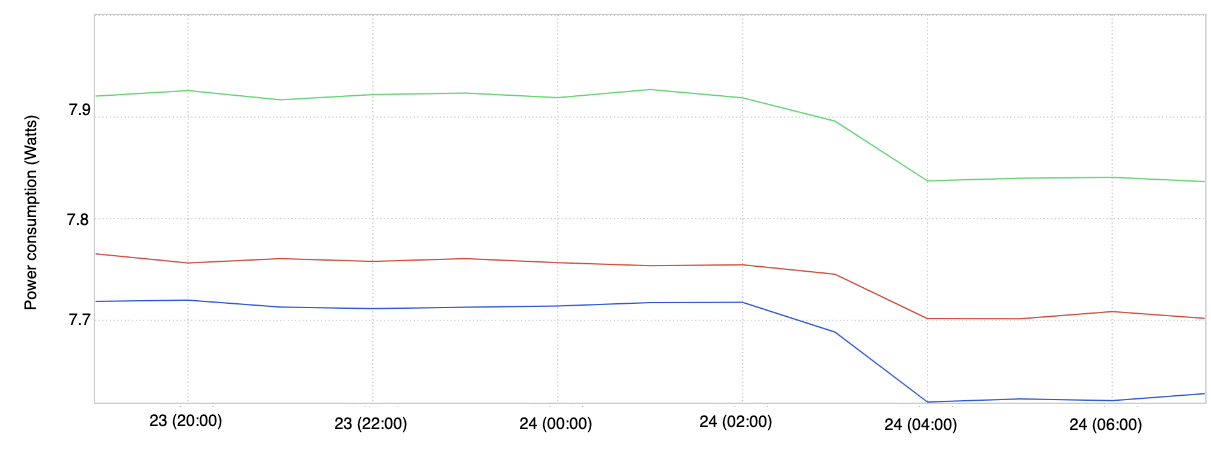}
    \caption{SSD Power consumption - average}
    \label{fig:power_drop}
\end{figure}

Figure \ref{fig:power_drop} shows the change in SSD power consumption associated with the same fan policy change. The three lines in the graph plot the average power consumption for SSDs in three different LinkedIn datacenters. 

\subsection{Internal Garbage Collection}
The SSD's FTL (Flash Translation Layer) is responsible for converting addresses sent by the driver into physical addresses corresponding to flash pages. In SSDs a page is the lowest resolution for writes. However, to write to a previously written-to page, it is necessary to first clear, or "erase" it. The resolution of erasure is an "Erase-Block". Each erase block is made up of multiple pages, each of which is cleared on erasing the block. Thus, each time a page must be erased, all other live pages in that block must be relocated before erasure takes place. This process is referred to as garbage collection.

\begin{figure}[ht]
    \centering
    \includegraphics[width=\columnwidth]{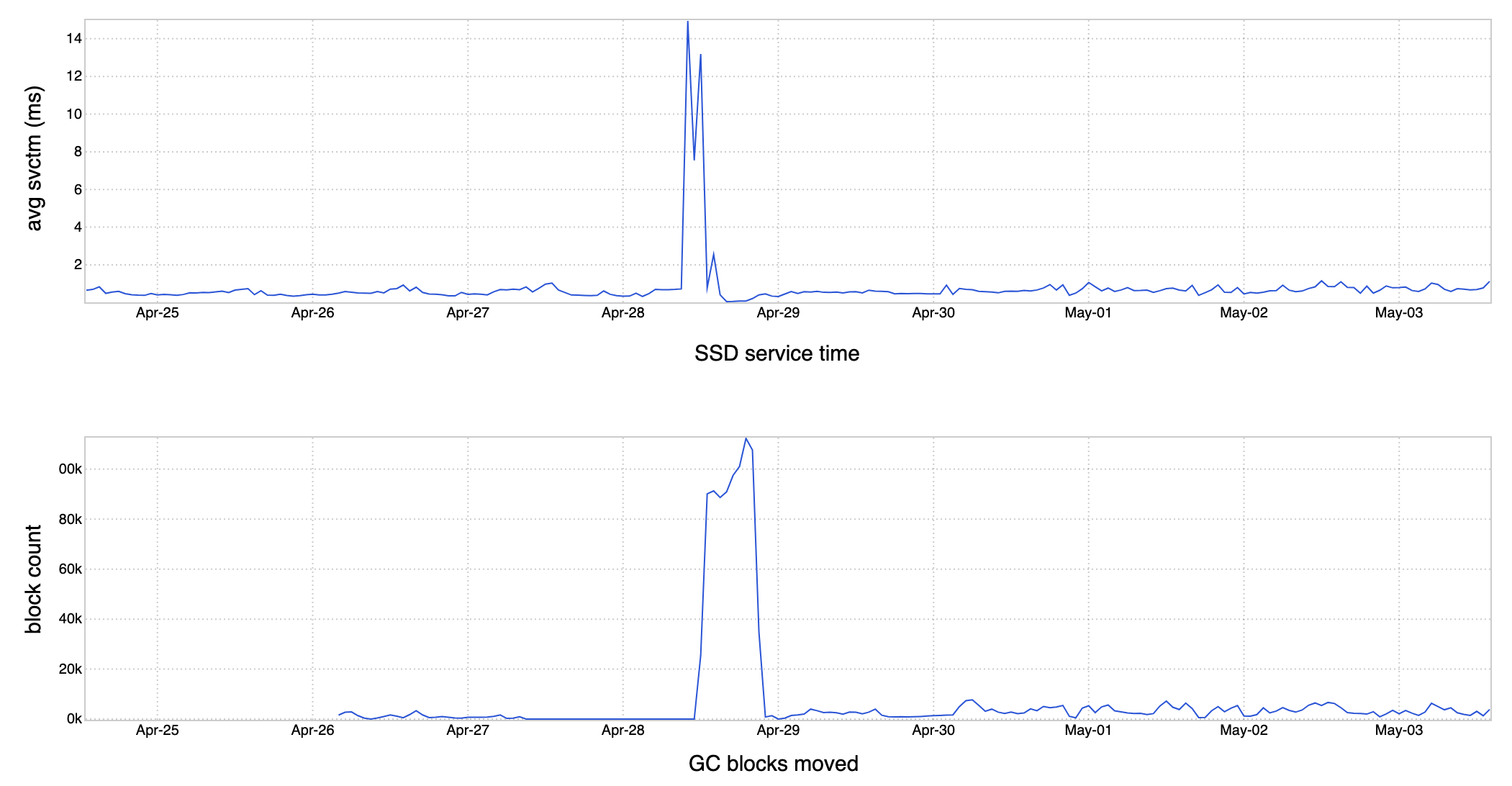}
    \caption{SSD IO service time and GC block movement}
    \label{fig:igc_svctm}
\end{figure}

In Virident SSDs, there are two modes of garbage collection:
\begin{enumerate}
    \item On-demand garbage collection
    \item Idle-time garbage collection
\end{enumerate}
On-demand garbage collection occurs only when a write command has been received for a cell that has already been written to, and there is no free page from which to service this write. On-demand GC thus incurs a time penalty on the IO activity which triggered it and all IO operations which occur while relocation is in progress. Idle-time GC, meanwhile performs garbage collection when the SSD is idle, and not serving several IO requests. In the default configuration used at LinkedIn, the on-demand GC mode is enabled.
We were able to narrow this down as the cause of several IO latency issues which impacted Espresso, LinkedIn's distributed document store \cite{qiao2013brewing}. When a drive is written to for a large period of time, without sufficient erasures taking place, IO to the device becomes slower, as a majority of writes involve on-demand GC taking place. This situation worsens until a large GC event takes place, recovering enough blocks to efficiently serve future IO requests. 
During this time, IO latency spikes significantly. This exact scenario occurred in a live Espresso storage node, causing the application's latency to spike beyond SLA. This can be observed in Figure \ref{fig:igc_svctm}. The graph plots IO service time (read at OS-level) and GC activity, measured as block movement, read from the SSD driver. Here, we can see that IO latency increased significantly in the period just before we see a spike in GC activity. As the GC activity continued, the IO latency dropped once there were enough free pages to service the incoming writes.
Based on these observation, a recommendation was made that all newly deployed PCIe SSDs be configured with idle-time GC enabled. Since this change, there has not been a single case where high latency was attributed to SSD GC activity.

Note that a functionality called TRIM exists for SSDs, which notifies the drive that a particular page is no longer required by the filesystem. The drive can then reclaim this page to serve future writes. However, our Virident SSDs do not support this command. A similar chain of events as described above was also observed in an NVMe SSD, for a workload which collects and persists system logs from all hosts in a fabric. This workload has consistently high throughput, owing to the volume of log files written each minute. The host also sees very high drive space utilization; upto 98\% filesystem utilization in production. In one instance, the filesystem utilization reached 99\%, which caused IO latency to spike as the SSD had limited overprovisioned space to service these writes. This latency grew, until the drive was full, and stopped servicing IO operations entirely. The metrics described previously allowed operations teams to quickly identify the root cause of this issue and trigger an FTL reset. This cleared all data, reset the drive to a fresh state, and restored IO performance to expected levels. Previously, such an issue would cause the drive to be replaced. However, with present monitoring and understanding of root cause, the issue was debugged and fixed within 10 minutes of the failure occurring.

\section{Impact}
As alluded to in Section 1, a single SSD failure results in downtime for the application, time from systems operations and datacentre technicians in case of replacement, and also has a monetary impact. In this section we describe summary statistics describing the impact of our work, and present some specific examples where our data was able to support rigorous root cause analysis for faults, as well as predict certain classes of errors.

\subsection{Outages}
\begin{figure}[ht]
    \centering
    \includegraphics[width=\columnwidth]{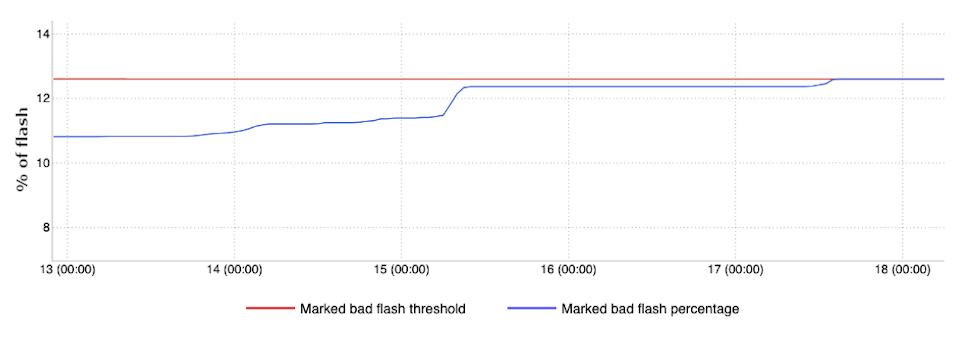}
    \caption{Bad flash}
    \label{fig:badflash}
\end{figure}

For Virident SSDs, an important metric mentioned previously was $Marked Bad Flash$, This refers to the percentage of the drive's total flash storage that has been marked unusable by the drive. A flash block may be marked bad for the following reasons:
\begin{enumerate}
    \item Write failure count crosses a threshold
    \item Erase failure count crosses a threshold
    \item Correctable read error count crosses a threshold
    \item Uncorrectable read error count crosses a threshold
\end{enumerate}
These thresholds are internal to the drive, and are not exposed by the driver. There is also a threshold for the amount of flash on the SSD that is allowed to go bad before the SSD goes into "READ-ONLY" mode, and stops accepting write commands. This event occurring while serving traffic is a critical error which immediately impacts the application. Both the percentage of bad flash and the threshold for the same are metrics exposed by the driver, and plotted in InGraphs by our monitoring applications. By observing the difference between these two graphs, it is possible to alert when a drive is close to failure. Another event which causes several flash blocks to be marked bad consecutively is when the driver is marking an entire subchannel in the SSD bad. This occurs when several blocks in a subchannel have been marked bad, and the entire subchannel is deemed to be unusable. When a subchannel is marked bad, all the remaining blocks in it are also marked bad, and any data written to them is moved to another subchannel. Thus, when a single subchannel is being marked bad, it results in several blocks being marked bad in rapid succession. When a subchannel is marked bad, it typically results in an increase in bad flash percentage of between 1 and 2 points. Using this knowledge, systems operations teams are now able to alert application SREs when an SSD is about to fail under such circumstances. This allows the SRE's to fail the application out of the affected node, without suffering from IO errors. An example of such a scenario can be seen in Figure \ref{fig:badflash}.
During an audit of all PCIe drives, it was observed that one host's SSD was rapidly approaching the threshold. The MySQL application running on this host was stopped and failed over by the SREs. This caused IO on the drive to be stopped entirely, resulting in there being no further increase in bad flash percentage. This can be seen in Figure \ref{fig:badflash} on the 15th. On the 17th, a synthetic load was run on the SSD using FIO \cite{axboe2019}. During this, the drive threshold was breached, and resulted in the drive going into READ-ONLY mode, hence confirming the original hypothesis. 

As a result of this case, an automatic alert has been created to detect such events. Through this system, 5 predictive alerts have been raised in a period of 2 months. The lead time for these alerts varies between 30 minutes to 48 hours, depending on the workload running on the drive and the drive capacity.

\subsection{Man hours}
\begin{figure}[ht]
    \centering
    \begin{minipage}{0.45\textwidth}
        \centering
        \includegraphics[width=0.9\textwidth]{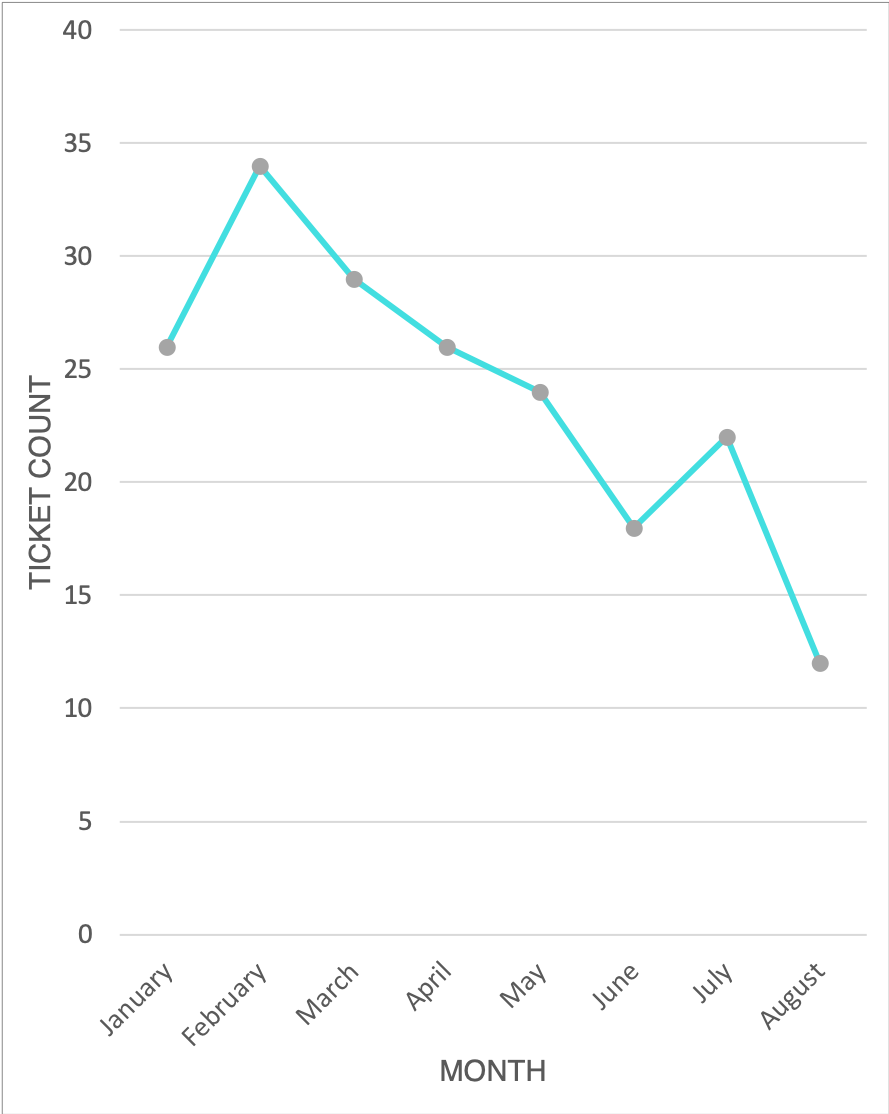}
        \caption{Thermal and Power throttling ticket load - 2019}
        \label{fig:throttle_ticks}
    \end{minipage}\hfill
    \begin{minipage}{0.45\textwidth}
        \centering
        \includegraphics[width=0.9\textwidth]{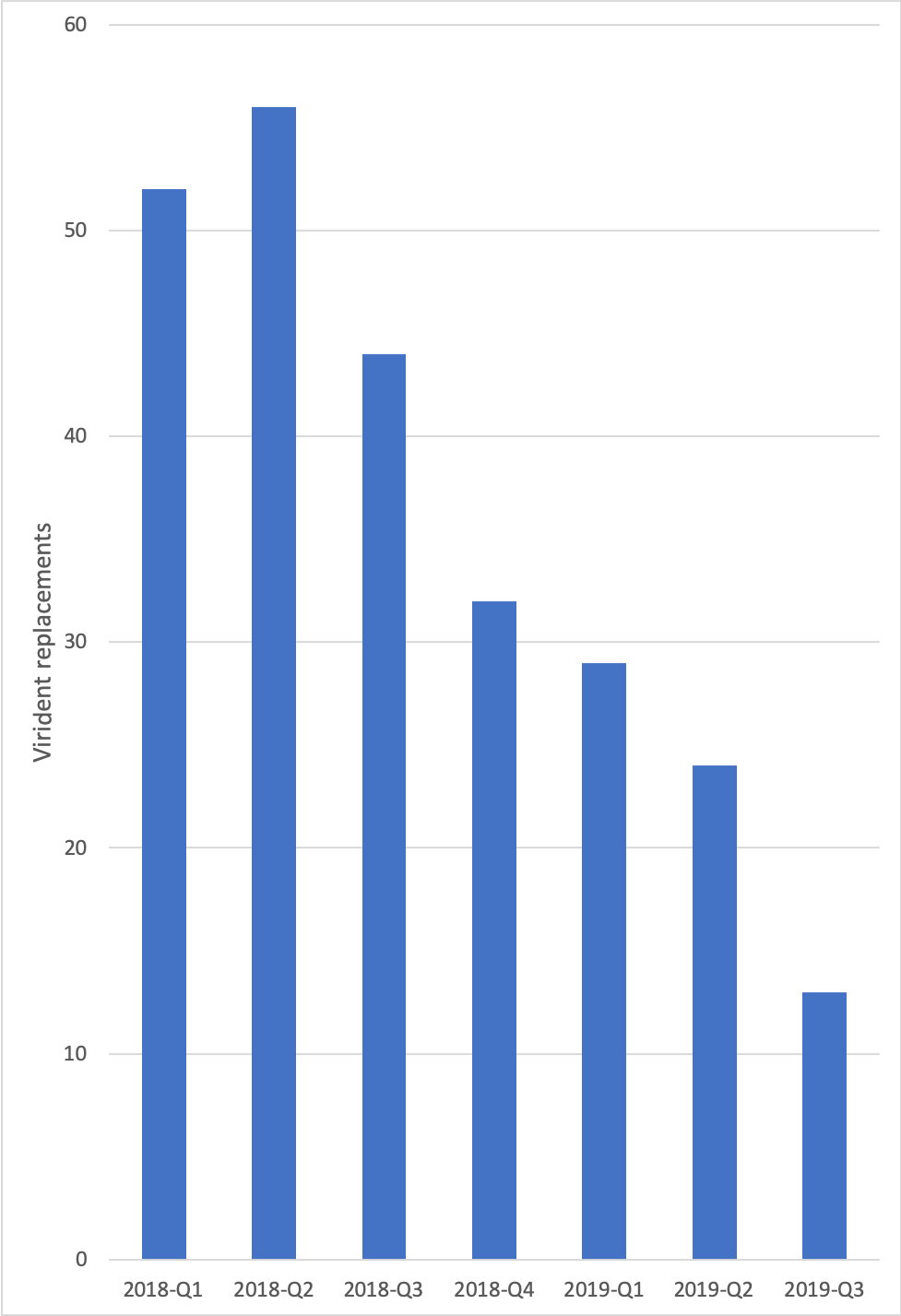}
        \caption{PCIe SSD Replacements}
        \label{fig:pcie_replacements}
    \end{minipage}
\end{figure}

As was described in Section \ref{ssdsatli}, PCIe SSDs contribute 77\% of the SSD related fault tickets that are raised for systems operations. What is not apparent here is that  each ticket typically takes a very large amount of time to resolve. Each SSD ticket involves contacting the application owner to schedule downtime for the application, investigating logs and metrics for the root cause of failure, and finally proposing and executing a solution. In total, each ticket costs multiple hours of an operations engineer's time. One of the most common classes of tickets is tickets related to thermal and power throttling. When the drive detects power or temperature crossing internal thresholds, it reduces its performance in order to draw less power from the bus and dissipate less heat. In Figure \ref{fig:throttle_ticks} we show the trend in tickets related to thermal and power throttling.

The number of throttling tickets for PCIe SSDs is a near-monotonic decreasing trend. This is attributed to the proactive fan policy change described in Section \ref{obs}. Graphs from the section show a drop in operating temperatures and power draw on the bus

\subsection{Financial impact}
In Figure \ref{fig:pcie_replacements} we present the number of PCIe drives replaced every quarter. The monotonically decreasing trend clearly signals significant improvement in the ability to provide effective diagnosis and fixes for SSD faults. We can compare the number of drives replaced in the first 3 quarters of 2019, which is when our project was in effect, with the first three quarters of 2018, which will act as a baseline. In Q1-Q3 of 2019, 66 drives were replaced. This is a much smaller than the 152 cards which were replaced during the same time period in 2019. We attribute this reduction primarily to the work described in this paper, since no other configuration or inventory change was made during this period.
When a PCIe SSD is to be removed, it is replaced with an NVMe SSD. For this, the host's chassis must be replaced, since the older chassis are not NVMe compatible. This means that the cost of replacing a PCIe SSD is the combination of the cost of the new drive and the cost of the new chassis. Through some computation on the number of drive replacements reduced and the respective costs of the NVMe SSDs and new chassis, we have quantified the financial impact of our work to be $\$300,000$ over a period of 9 months.

\section{Related work}

The key focus of our literature survey was to identify the metrics that are best representative of errors and misconfiguration in SSDs.
Several previous studies of HDD (Spinning disk) failure trends exist in literature. Pinheiro et. al \cite{pinheiro2007failure} describe the trend in several SMART metrics in failing drives in Google's datacenters. They identify some specific metrics as being prevalent in a majority of failing drives, but find that none of the studied metrics is a good candidate to build a predictive system.
Zhu et. al \cite{zhu2013proactive} describe a system for using SMART attributes to predict hard drive failure. The paper describes a system which is able to predict drive failures with an accuracy upwards of 95\%, using just 10 SMART attributes. The data used for this paper's experimentation comes from Baidu's datacentres.
Xu et. al \cite{xu2018improving} describe a prediction model for spinning disks to improve cloud service reliability. The paper describes a variety of metrics divided into SMART metrics describing hardware parameters, and system metrics describing operating system and file system related attributes.
A detailed study of SSD failures can be found in Meza et. al \cite{meza2015large}. This study provides an insightful discussion of the architecture of an enterprise SSD, and describes the points of failure. Correlation between several factors and error rates are identified.

\section{Conclusion}
In this paper we have presented the design of the SSD monitoring infrastructure deployed in LinkedIn's datacenters. We have described the distribution of SSDs, typical workloads and typical error classes. We have described the metrics that are being actively monitored, and the infrastructure for collecting and plotting the same.
We have covered some faults and misconfigurations that were possible to observe through the new system, and the actions taken to remedy the same. 
We have described how the impact of our work extends beyond a few specific cases to a broad improvement in reliability. We have also presented a quantification of the monetary impact of our work. It is our hope that the insights we have gained from our work have demonstrated the advantage of monitoring Solid State Drives to optimize lifecycle management.

\printbibliography

\end{document}